\pdfoutput=1
\documentclass[preprint,3p,times]{elsarticle}

\usepackage{amsmath}
\usepackage{amssymb}
\biboptions{sort&compress}

\usepackage[figuresright]{rotating}

\usepackage{graphicx}

\graphicspath{{./}}
\DeclareGraphicsExtensions{.png}

\begin{document}

\begin{frontmatter}

%% Title, authors and addresses

%% use the tnoteref command within \title for footnotes;
%% use the tnotetext command for the associated footnote;
%% use the fnref command within \author or \address for footnotes;
%% use the fntext command for the associated footnote;
%% use the corref command within \author for corresponding author footnotes;
%% use the cortext command for the associated footnote;
%% use the ead command for the email address,
%% and the form \ead[url] for the home page:
%%
%% \title{Title\tnoteref{label1}}
%% \tnotetext[label1]{}
%% \author{Name\corref{cor1}\fnref{label2}}
%% \ead{email address}
%% \ead[url]{home page}
%% \fntext[label2]{}
%% \cortext[cor1]{}
%% \address{Address\fnref{label3}}
%% \fntext[label3]{}

% \dochead{}
%% Use \dochead if there is an article header, e.g. \dochead{Short communication}

% \title{Investigation of jet quenching and elliptic flow within a pQCD-based partonic transport model}
% \title{Jet and bulk physics within a pQCD-based partonic transport model}
\title{Jets and flow within a pQCD-based partonic transport model}

%% use optional labels to link authors explicitly to addresses:
%% \author[label1,label2]{<author name>}
%% \address[label1]{<address>}
%% \address[label2]{<address>}

\author{Oliver Fochler, Zhe Xu, Carsten Greiner}

\address{Institut f\"ur Theoretische Physik\\
Goethe-Universit\"at Frankfurt\\
Max-von-Laue-Str. 1, D-60438 Frankfurt am Main, Germany}

\begin{abstract}
We present fully dynamic simulations of heavy ion collisions at RHIC energies within the perturbative QCD-based partonic transport model BAMPS, focusing on the simultaneous investigation of jet-quenching and elliptic flow. The model consistently features elastic and inelastic $2 \leftrightarrow 3$ processes, the latter being based on the Gunion-Bertsch matrix element. We discuss first attempts to extend the model to include light quark degrees of freedom and study the energy loss of high energy gluons and quarks in a static partonic medium. The difference between gluons and quarks in inelastic processes is found to be weaker than expected from color factors, due to a self-quenching effect associated with a cut-off modeling the LPM effect.
\end{abstract}

\begin{keyword}
Hard Probes 2010 \sep Partonic transport model \sep Energy loss \sep Elliptic flow
%% keywords here, in the form: keyword \sep keyword

%% MSC codes here, in the form: \MSC code \sep code
%% or \MSC[2008] code \sep code (2000 is the default)

\end{keyword}

\end{frontmatter}

%%
%% Start line numbering here if you want
%%
% \linenumbers

%% main text
\section{Introduction}
\label{sec:introduction}

Two of the most important phenomena observed in collisions of heavy nuclei at the Relativistic Heavy Ion Collider (RHIC) go by the names of \emph{jet quenching} and \emph{elliptic flow}.

The RHIC experiments have established that particles with high transverse momenta are suppressed in heavy ion collisions with respect to a scaled proton-proton reference \cite{Adler:2002xw,Adcox:2001jp}. This quenching of jets is commonly attributed to energy loss on the partonic level. It could therefore provide means of investigating the properties of the medium, the quark--gluon plasma (QGP), that causes the modifications of the high-$p_{T}$ particles as they traverse it after their production in initial hard processes. Jet quenching on the single hadron spectra level is usually quantified in terms of the nuclear modification factor $R_{AA}$.
% \begin{equation} \label{eq:RAA}
% R_{AA}=\frac{d^{2}N_{AA}/dp_{T}dy}{T_{AA}d^{2}\sigma_{NN}/dp_{T}dy}
% \, .
% \end{equation}

The collective flow of the created medium \cite{Adler:2003kt, Adams:2003am} is usually quantified in terms of the coefficient $v_{2}$ in a Fourier expansion of the angular distribution $dN/d\phi$ and in this context often referred to as elliptic flow. The comparison of hydrodynamic calculations to data indicates that the viscosity of the QGP is quite small \cite{Romatschke:2007mq}, possibly close to the conjectured lower bound $\eta/s = 1 / (4 \pi)$ from an AdS/CFT correspondence \cite{Kovtun:2004de}.

The energy loss of partonic jets on the other hand can be treated in terms of perturbative QCD (pQCD) and most theoretical schemes attribute the main contribution to partonic energy loss to radiative processes \cite{Baier:1998yf,Gyulassy:2000er,Jeon:2003gi,Wicks:2005gt}.

It is a major challenge to combine high-$p_{T}$ physics and bulk evolution within a common framework. Recently, efforts combining pQCD-based energy loss calculations with hydrodynamic modeling of the medium have been intensified, using results from hydrodynamical simulations as an input for the medium evolution in jet--quenching calculations (see \cite{Bass:2008rv} for an overview). However, these approaches still treat medium physics and jet physics in the QGP on different grounds. No schemes are available that cover the full dynamics of the interplay between jets and the medium, i.e. that consistently include modifications of the medium caused by the traversing jets.

Partonic transport models might provide means to investigate bulk properties of the QGP and high-energy parton jets within a common physical framework automatically including the full dynamics of the evolution of the system. In previous publications \cite{Fochler:2008ts, Fochler:2010wn} we have explored the capabilities of the transport model BAMPS (\emph{Boltzmann Approach to Multi-Parton Scatterings}) with this goal in mind.

\section{The transport model}
\label{sec:BAMPS}

The microscopic transport model BAMPS \cite{Xu:2004mz,Xu:2007aa} is a tool to simulate the evolution of the QGP stage of heavy ion collisions. Partons are considered as massless Boltzmann particles whose interactions are based on leading order pQCD matrix elements and consistently include creation ($2 \rightarrow 3$) and annihilation ($3 \rightarrow 2$) processes. The test particle method is introduced to reduce statistical fluctuations.

BAMPS has so far been restricted to gluonic degrees of freedom, that already allows for the investigation of crucial features of the medium evolution in heavy ion collisions. Nevertheless, the contribution of light quarks is of great interest and we will discuss a possible extension of BAMPS to quark degrees of freedoms.

For elastic interactions Debye screened cross section in small angle approximation are used, for example
\begin{align}
 \frac{d\sigma_{gg\to gg}}{dq_{\perp}^2} &= \frac{9\pi\alpha_{s}^{2}}{(q_{\perp}^2+m_D^2)^2} 
&& \text{or}
 &\frac{d\sigma_{qg \rightarrow qg}}{dq_{\perp}^{2}} &= \frac{2\pi \alpha_{s}^{2}}{(q_{\perp}^{2} + m_{D}^{2} )^{2}} \, .
\end{align}
The Debye screening mass is computed via $ m_{D}^{2} = d_G \pi \alpha_{s} \int \frac{d^{3}p}{(2\pi)^{3}} \frac{1}{p} (N_{c}f_{g} + N_{f}f_{q} )$ from the local gluon and quark distributions, $f_{g}$ and $f_{q}$, where $d_G = 16$ is the gluon degeneracy factor for $N_c = 3$.

\begin{figure}[tbh]
  \centering
  \begin{minipage}[t]{0.4\textwidth}
    \includegraphics[width=\linewidth]{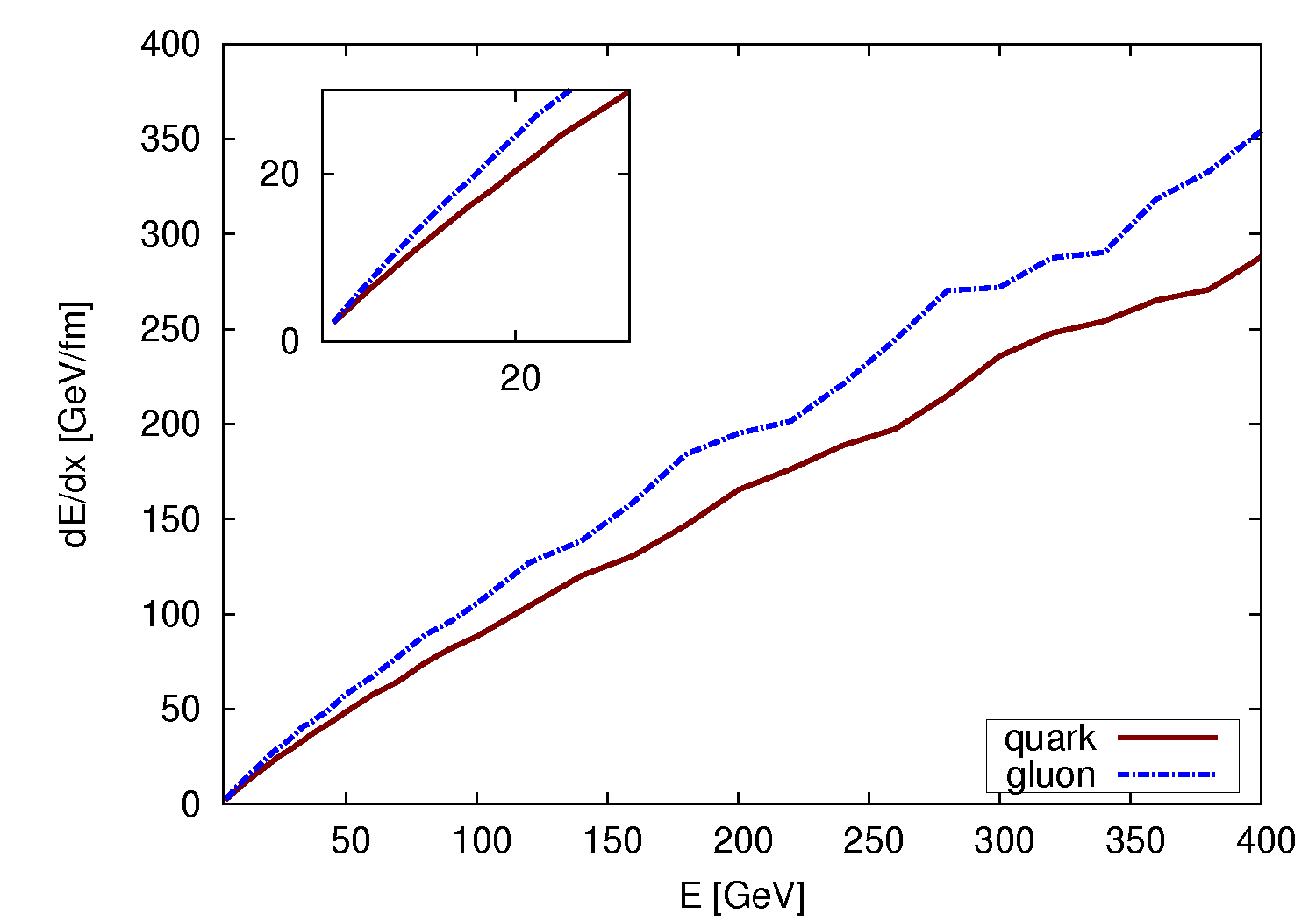}
  \end{minipage}
  \begin{minipage}[t]{0.4\textwidth}
    \includegraphics[width=\linewidth]{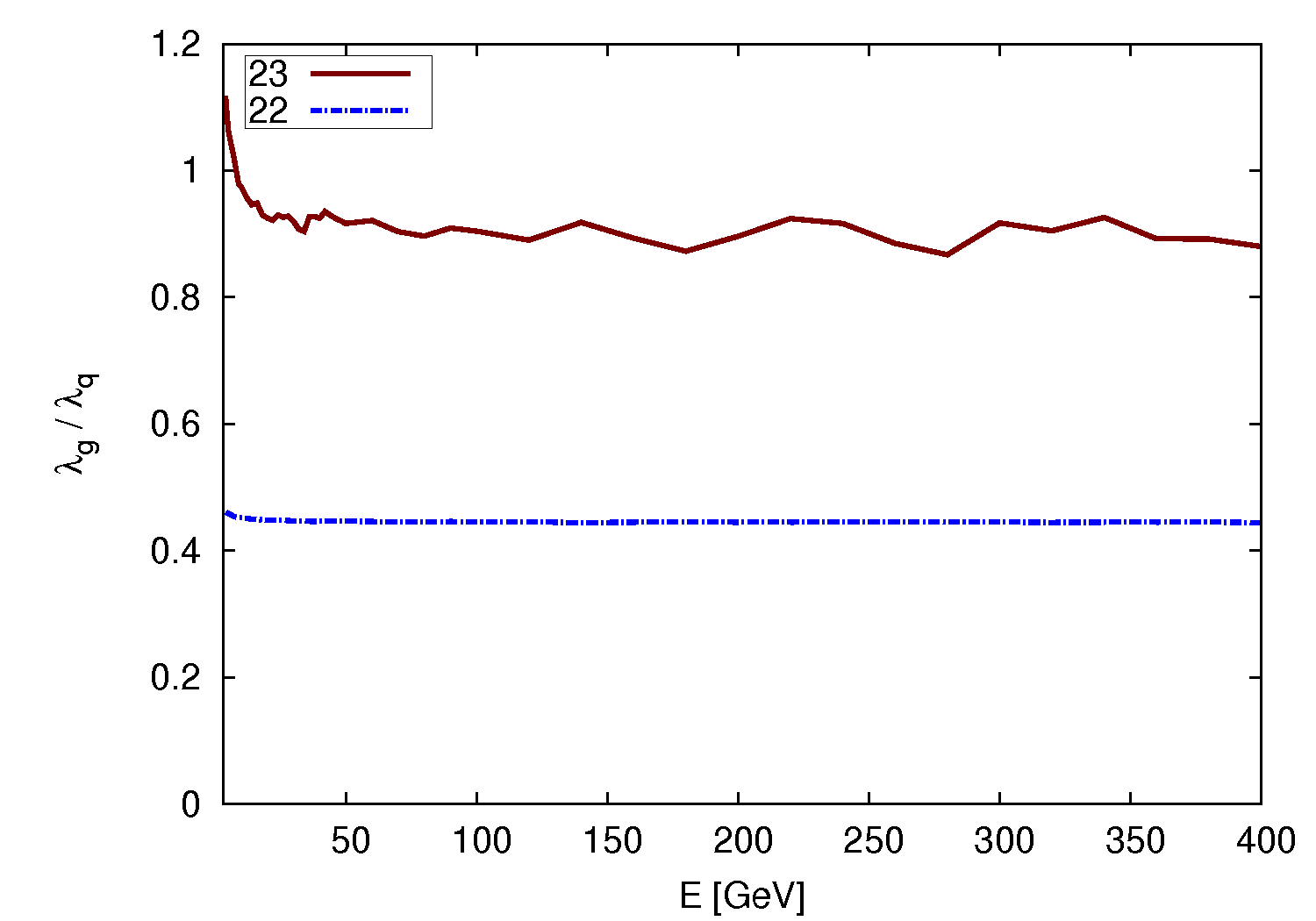}
  \end{minipage}
  \caption{Left panel: Total $dE/dx$ for a high energy gluon and a high energy light quark as a function of energy. Medium temperature $T=0.4\,\mathrm{GeV}$.\newline Right panel: Ratio of the mean free path for gluons to the mean free path for light quarks for purely $2\rightarrow2$ interactions and for $2\rightarrow 3$ processes.}
\label{fig:dEdx}
\end{figure}

Inelastic processes are treated via the Gunion-Bertsch matrix element \cite{Gunion:1981qs}. For $gg\rightarrow ggg$ it reads:
\begin{equation} \label{eq:gg_to_ggg}
\left|\mathcal{M}_{gg \to ggg}\right|^2 = \frac{72 \pi^2 \alpha_s^2 s^2}{(\mathbf{q}_{\perp}^2+m_D^2)^2}\,
 \frac{48 \pi \alpha_s \mathbf{q}_{\perp}^2}{\mathbf{k}_{\perp}^2 [(\mathbf{k}_{\perp}-\mathbf{q}_{\perp})^2+m_D^2]}
\Theta\left( \Lambda_g - \tau \right)
\text{,}
\end{equation}
where $\mathbf{q}_{\perp}$ and $\mathbf{k}_{\perp}$ denote the perpendicular components of the momentum transfer and of the radiated gluon momentum in the center of momentum (CM) frame of the colliding particles, respectively. The principle of detailed balance yields the matrix element for the back reaction. Inelastic processes involving quarks, e.g. $qq \rightarrow qqg$, are dealt with by considering that the $2 \rightarrow 3$ matrix element factorizes into a probability for the emission of a gluon and an elastic contribution. This in mind, the matrix element for e.g. $qq \rightarrow qqg$ can be obtained from (\ref{eq:gg_to_ggg}) via $\left|\mathcal{M}_{qq \to qqg}\right|^2 = \frac{\left|\mathcal{M}_{qq \to qq}\right|^2}{\left|\mathcal{M}_{gg \to gg}\right|^2}\left|\mathcal{M}_{gg \to ggg}\right|^2$. Taking the small angle approximation for the elastic matrix elements, the prefactor $\left|\mathcal{M}_{qq \to qq}\right|^2 / \left|\mathcal{M}_{gg \to gg}\right|^2$ simply reduces to a momentum independent color factor. 

The Theta function in (\ref{eq:gg_to_ggg}) is an effective implementation of the LPM (Landau, Pomeranchuk, Migdal) effect \cite{Migdal:1956tc}, that describes coherence effects in multiple bremsstrahlung processes. This interference effect cannot be incorporated directly into a semi-classical microscopic transport model such as BAMPS, so the cut-off $\Theta\left( \Lambda - \tau \right)$ ensures that successive $2 \rightarrow 3$ processes are independent of each other. $\tau$ is the formation time of the gluon emitted with transverse momentum $k_{\perp}$ and $\Lambda$ denotes the mean free path, i.e. the time between successive interactions, of the parent parton. When comparing the formation time to the mean free path of the parent parton, attention needs to be paid to the reference frames. This renders the cut-off dependent on the boost $\vec{\beta}$ between the plasma rest frame and the center of momentum (CM) frame in which (\ref{eq:gg_to_ggg}) is evaluated \cite{Fochler:2010wn} and numerically further complicates the calculations.

\section{Results}
\label{sec:results}

Studying the evolution of high energy partons in a static thermal medium with fixed temperature, we find that within our setup the mean energy loss per unit path length $dE/dx$ is strongly dominated by inelastic $2 \rightarrow 3$ processes. $dE/dx$ in a static medium is calculated from the average energy loss per collision and the average interaction rate as $dE/dx = dE/d(ct) = \sum_{i} \langle \Delta E^{i} \rangle R^{i}$, where the sum runs over all possible interaction processes.

While $dE/dx$ from binary processes follows the well know $T^{2} \ln(ET)$ dependence and is moderate at $dE/dx_{gg\rightarrow gg} \approx 1.2\,\mathrm{GeV}/\mathrm{fm}$ for $E=50\,\mathrm{GeV}$ and $T=0.4\,\mathrm{GeV}$, the energy loss from interactions according to the Gunion-Bertsch matrix element rises roughly linearly with $E$ and is rather strong with $dE/dx_{gg\rightarrow ggg} \approx 32.6\,\mathrm{GeV}/\mathrm{fm}$ for the same medium parameters \cite{Fochler:2010wn}. The total $dE/dx$ can be seen in the left-hand panel of fig. \ref{fig:dEdx}.

For elastic processes the mean free path and $dE/dx$ of a high energy gluon and a light quark differ by roughly a factor of $2$ as expected from the color factors. For $2\rightarrow 3$ processes however, the difference is much weaker. Only about $20\%$ in the energy loss $dE/dx$ and the mean free paths, see fig. \ref{fig:dEdx}. This is due to the cut-off $\Theta\left( \Lambda - \tau \right)$ that basically self-quenches the difference in the matrix elements for gluons and quarks. When the mean free path is larger, as is expected for quarks from the color factors, the step function for the integration over the matrix element (\ref{eq:gg_to_ggg}) becomes less restrictive, yielding a larger cross section and thus reducing the mean free path. The results shown in fig. \ref{fig:dEdx} are self-consistently obtained from iterative calculations of the mean free path $\lambda$.

\begin{figure}[tbh]
  \centering
  \begin{minipage}[t]{0.4\textwidth}
    \includegraphics[width=\linewidth]{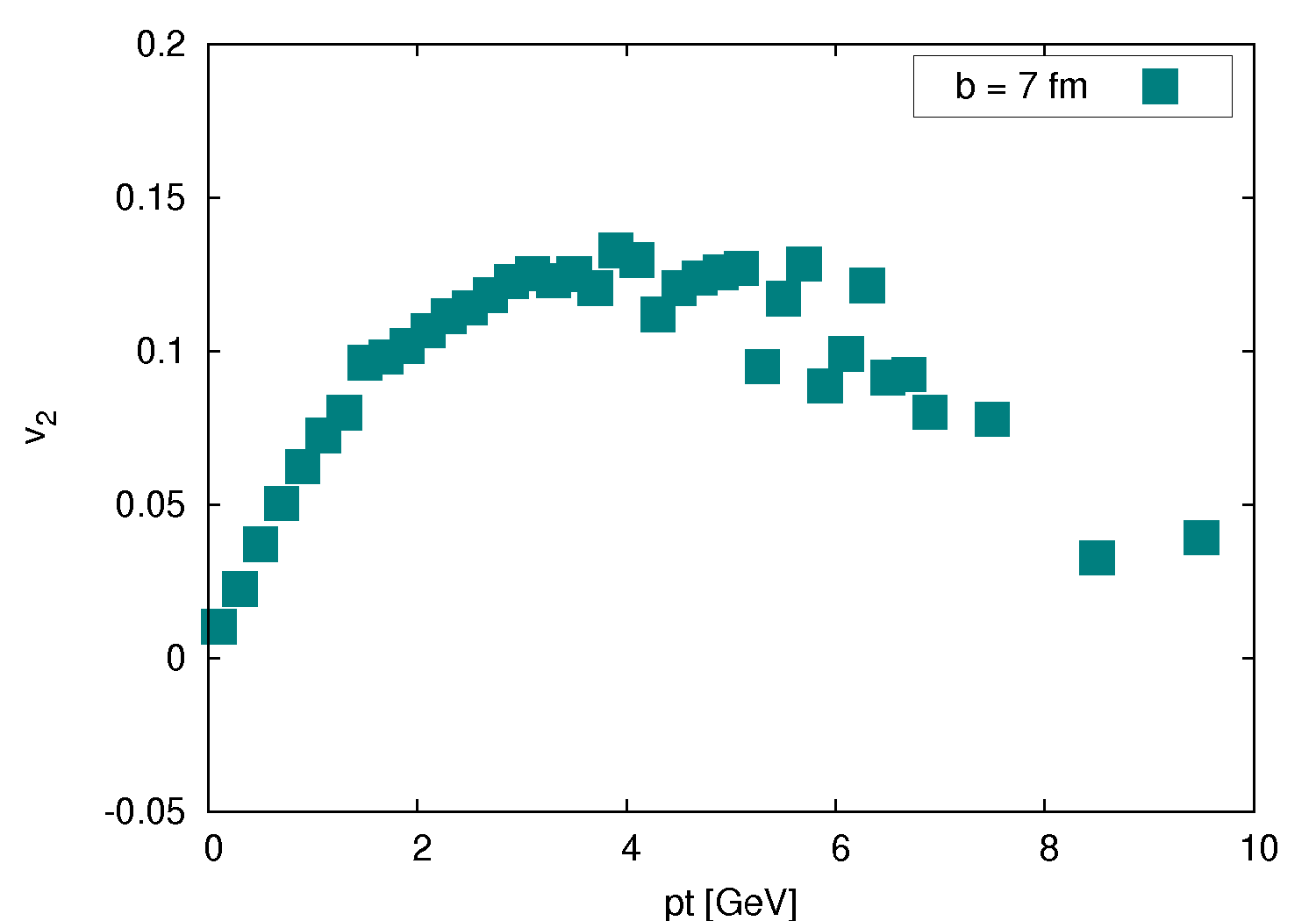}
  \end{minipage}
  \begin{minipage}[t]{0.4\textwidth}
    \includegraphics[width=\linewidth]{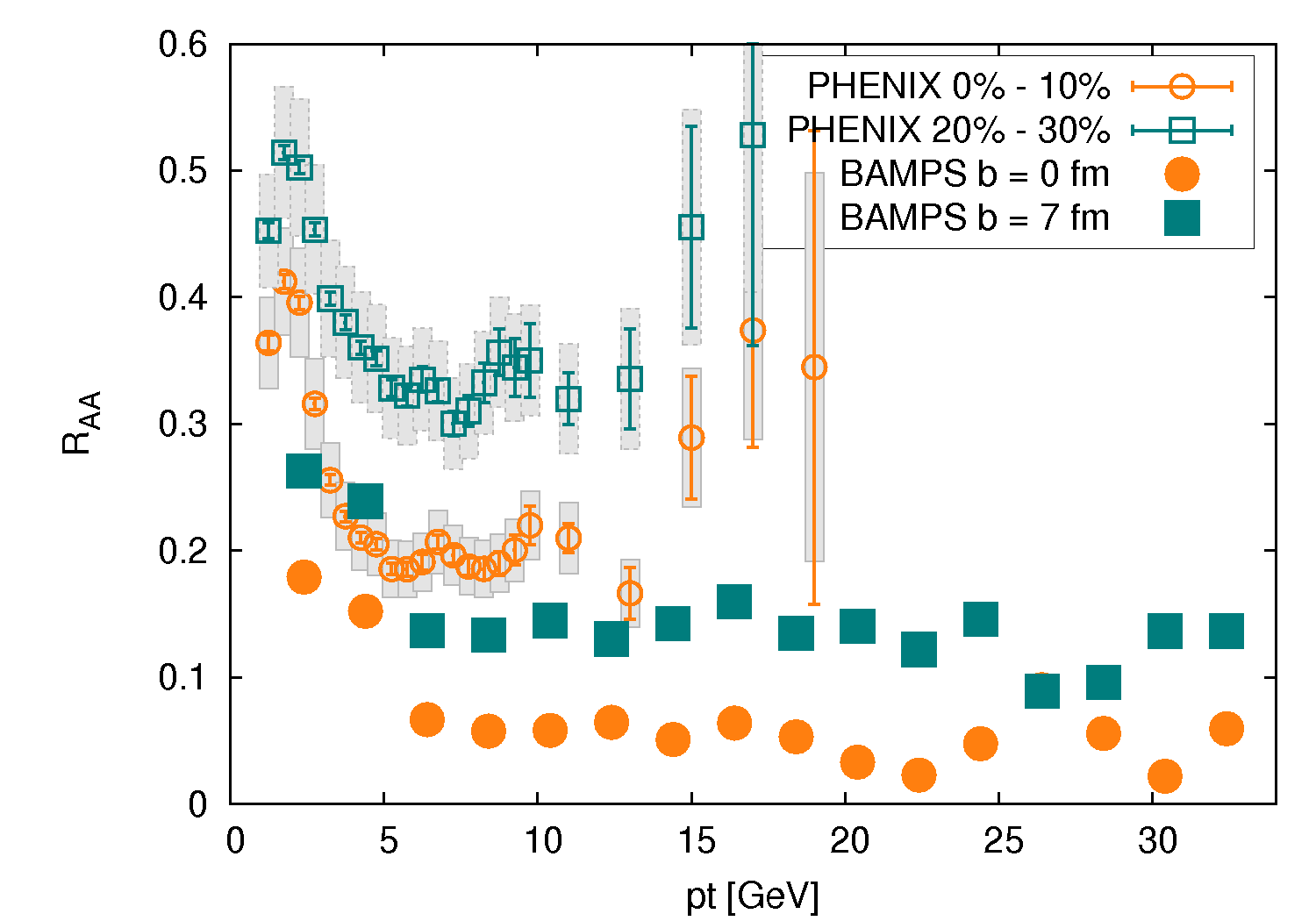}
  \end{minipage}
  \caption{Left panel: Elliptic flow $v_{2}$ for gluons in simulated Au~+~Au collisions at 200 AGeV with $b=7\,\mathrm{fm}$. \newline
  Right panel: Gluonic $R_{AA}$ as extracted from BAMPS simulations for $b=0\,\mathrm{fm}$ and $b=7\,\mathrm{fm}$. For comparison experimental results from PHENIX \cite{Adare:2008qa} for $\pi^{0}$ are shown for central ($0 \%$ - $10 \%$) and off-central ($20 \%$ - $30 \%$) collisions. See \cite{Fochler:2010wn} for details.}
\label{fig:v2RAA}
\end{figure}

Since the difference in the differential energy loss in static media from our treatment of quark inelastic interactions is rather weak, only a small difference in the nuclear modification factors of gluons and light quarks is to be expected in fully dynamic simulations of Au~+~Au collisions at RHIC energies. However, these calculations have not been fully completed yet and will be presented in a later publication.

For a purely gluonic medium $R_{AA}$ in central Au~+~Au collisions (initial gluon distribution from a Glauber plus mini-jet model) is flat at roughly $R_{AA}^{\mathrm{gluons}} \approx 0.053$ and in reasonable agreement with analytic results from Wicks et al. \cite{Wicks:2005gt, Fochler:2008ts}, though the suppression of gluon jets in BAMPS appears to be slightly stronger. In non-central collisions the suppression is reduced but $R_{AA}$ remains flat. The gluon $R_{AA}$ at $b=7\,\mathrm{fm}$ increases by a factor of roughly $2.5$ compared to $b=0\,\mathrm{fm}$. The relative increase is less pronounced ($\sim 1.7$) in the pion data, comparing $0 \%$-$10 \%$ and $20 \%$-$30 \%$ central collisions. See right panel of fig. \ref{fig:v2RAA}.
The $v_{2}(p_{T})$ of high-$p_{T}$ gluons at $b=7\,\mathrm{fm}$ is rising up to $v_{2} \approx 0.12$ at $p_{T} \approx 4\,\mathrm{GeV}$ and slightly decreases afterwards (left panel of fig. \ref{fig:v2RAA}). This behavior is in good qualitative agreement with recent RHIC data \cite{Abelev:2008ed}.

\section{Summary}
\label{sec:summary}

The transport model BAMPS can be used to study the energy loss of high-$p_{T}$ partons and the elliptic flow within a common framework. We find gluonic $R_{AA}$ for central Au~+~Au to be in reasonable agreement with analytic results and the integrated elliptic flow can be reproduced nicely \cite{Xu:2007jv}. Inelastic processes based on the Gunion-Bertsch matrix element strongly dominate the energy loss of high-$p_{T}$ partons.

First attempts to extend the model to include light quarks show that the energy loss of quarks and gluons differs by less than the expected color factor due to a self-quenching effect caused by the effective implementation of the LPM suppression. This will be studied in more detail in upcoming works, together with the application of fragmentation functions to the high--$p_{T}$ sector allowing for more direct comparison to hadronic observables. Also the simulation of LHC energies and a careful scan of various impact parameters is on the agenda, improving the comparability to experimental data. 

\section*{Acknowledgments}
This work has been supported by the Helmholtz International Center for FAIR within the framework of the LOEWE program launched by the State of Hesse. The simulations have been performed at the Center for Scientific Computing (CSC) at the Goethe University Frankfurt.

%% The Appendices part is started with the command \appendix;
%% appendix sections are then done as normal sections
%% \appendix

%% \section{}
%% \label{}

%% References
%%
%% Following citation commands can be used in the body text:
%% Usage of \cite is as follows:
%%   \cite{key}         ==>>  [#]
%%   \cite[chap. 2]{key} ==>> [#, chap. 2]
%%

%% References with BibTeX database:

\bibliographystyle{elsarticle-num}
\bibliography{fochler.bib}

%% Authors are advised to use a BibTeX database file for their reference list.
%% The provided style file elsarticle-num.bst formats references in the required Procedia style

%% For references without a BibTeX database:

% \begin{thebibliography}{00}

%% \bibitem must have the following form:
%%   \bibitem{key}...
%%

% \bibitem{}

% \end{thebibliography}

\end{document}